\documentclass[12pt,useAMS,useAASTEX,usenatbib]{aastex}
\usepackage{emulateapj5}

\makeatletter

\newenvironment{apjemufigure}{%
\def\@captype{figure}%
\noindent\begin{minipage}{0.999\linewidth}\begin{center}}
{\end{center}\end{minipage}}
\makeatother

\def\healpix{H{\sc ealpix }}
\def\meanchi{\overline{\chi^2}}
\def\brameanchi{\langle \meanchi \rangle}  
\def\wmap{\hbox{\sl WMAP~}}

\def\etal{et al.}

\def\alm{a_{\ell m}}
\def\Ylm{Y_{\ell m}}
\def\Cl{C_{\ell}}

\def\summ{\sum_{m=-\ell}^{\ell}}
\def\suml{\sum_{\ell=0}^{\infty}}

\def\planck{{\it Planck }}

\newcommand{\tac}{{Theoretical Astrophysics Center, Juliane Maries Vej
30, DK-2100,  Copenhagen, Denmark}}
\newcommand{\nbi}{{Niels Bohr Institute, Blegdamsvej 17,
DK-2100 Copenhagen, Denmark}}
\newcommand{\sao}{{Special Astrophysical Observatory, Nizhnij Arkhyz,
Karachaj-Cherkesia, 369167, Russia}}
\newcommand{\nasa}{{NASA Ames Research Center, Space Science
Division, Moffett Field, CA 94035}}

\slugcomment{Submitted to {\it The Astrophysical Journal Letters}}

\begin{document}

\title{Non-Gaussianity of the derived maps from the first-year {\it WMAP} data}

\author{
Lung-Yih Chiang\altaffilmark{1},
Pavel D. Naselsky\altaffilmark{1,2},
Oleg V. Verkhodanov\altaffilmark{1,3},
Michael J. Way\altaffilmark{1,4}
}

\altaffiltext{1}{\tac}
\altaffiltext{2}{\nbi}
\altaffiltext{3}{\sao}
\altaffiltext{4}{\nasa}

\email{chiang@tac.dk}


\begin{abstract}
We present non-Gaussianity testing on recently-released derived
maps from the first-year \wmap data by Tegmark, de Oliveria-Costa and
Hamilton. Our test is based on a phase mapping technique which has
the advantage of testing non-Gaussianity at separate multipole
bands. We show that their foreground-cleaned map is against the
random-phase hypothesis at all 4 multipole bands centered around
$\ell=150$, 290, 400 and 500. Their Wiener-filtered
map, on the other hand, is Gaussian for $\ell
<250$, and marginally Gaussian for  $224 < \ell < 350$. However, we
see the evidence of non-Gaussianity for $\ell > 350$ as we detect certain
degrees of phase coupling, hence against the random-phase
hypothesis. Our phase mapping technique is particularly useful for
testing the accuracy of component separation methods. 
\end{abstract}

\section{Introduction}
With the first-year data release of the Wilkinson Microwave Anisotropy Probe
(\wmap), it has been proclaimed that we have entered the era of ``precision
cosmology''. The temperature fluctuations of the cosmic microwave
background (CMB) radiation are believed to be the imprint of 
primordial density fluctuations in the early Universe which give rise
to the large-scale structures we see today. Hence the data enable us
to test the statistical character of the primordial fluctuations, making
subsequent inferences on the topology and content of the
Universe. 
  
Although the \wmap team \citep{wmapng} claims that the signal is
Gaussian with 95\% confidence, the internal linear combination map released by the \wmap
team is not up for CMB studies due to ``complex noise properties'' \footnote{{\tt http://lambda.gsfc.nasa.gov/product/map/m\_products.html}}. Another
group led by M. Tegmark has performed an independent foreground
cleaning from the first-year \wmap data and made public their
whole-sky CMB maps. Their foreground-cleaned map (hereafter FCM) and the
Wiener-filtered map (WFM) are available online \citep{tegmarkweb}.

The FCM by the authors' definition is such that the
foreground contamination is removed as much as possible. As
foregrounds are rather non-Gaussian, any residual after cleaning would
manifest itself in the phase configuration. In this {\it Letter} we
display the phases of the FCM and the WFM with color coding and implement our
phase-mapping technique to test quantitatively the Gaussianity of both
maps, based on the random-phase hypothesis of
homogeneous and isotropic Gaussian random fields. Our phase mapping
technique can play a crucial role as a qualitative criterion for
component separation similar to the field of image reconstruction.  

\section{Gaussian Random Fields and the Random Phase Hypothesis}
The statistical characterization of temperature fluctuation of
CMB radiation on a sphere can be expressed as a sum over spherical
harmonics:
\begin{equation}
\Delta T(\theta,\varphi)=\suml \summ \alm \Ylm (\theta,\varphi),
\end{equation}
where $\alm=|\alm| \exp(i \phi_{\ell m})$. Homogeneous and isotropic Gaussian random fields (GRFs), as a result
of the simplest inflation paradigm, possess Fourier modes whose real
and imaginary parts are independently distributed. In other words,
they have phases $\phi_{\ell m}$ that are independently distributed
and uniformly random on the interval $[0,2\pi]$ \citep{bbks,be}. Thus
the spatial variations should constitute a {\it statistically
homogeneous and isotropic GRF} \citep{bbks} whose statistical properties are
completely specified by its angular power spectrum $\Cl$, 
\begin{equation}
\langle  a^{}_{\ell^{ } m^{ }} a^{*}_{\ell^{'} m^{'}} \rangle = \Cl \;
\delta_{\ell^{ } \ell^{'}} \delta_{m^{} m^{'}}.  
\end{equation}

The strict definition of a homogeneous and isotropic GRF
requires that the amplitudes are Rayleigh distributed and the phases
are random \citep{wc}. At the same time, the Central Limit Theorem guarantees
that a superposition of a large number of Fourier modes with random
phases will be Gaussian. Therefore the random-phase hypothesis on its own
serves as a definition of Gaussianity \citep{bbks}.

\section{Color-coded phase map of the derived \wmap maps}
\citet{tegmark} (TDH03) perform an
independent foreground analysis from the \wmap data and provide a FCM and WFM.
We first use a visual display of phases by colors to show phase associations \citep{color}. In color
image display devices, each pixel represents the intensity and color
at that position in the image. Two color schemes are usually used for
the quantitative specification of color, namely the Red-Green-Blue (RGB) and
Hue-Saturation-Brightness (HSB) color schemes. Hue is the term used to
distinguish between different basic colors (blue, yellow, red and so on).
Saturation refers to the purity of the color, defined by how much
white is mixed with it. Brightness indicates the overall intensity of
the pixel on a grey scale. The HSB color model is particularly useful
because of the properties of the `hue' parameter, which is defined as
a circular variable. Therefore we are mapping phases from 0 to $2\pi$
to the hue circle.

We have used the \healpix\footnote{\tt
http://www.eso.org/science/healpix/} package to produce $\alm$.
In Fig.~\ref{color} we show the color-coded phase gradient $D_\ell
\equiv \phi_{\ell+1,m}-\phi_{\ell,m}$ for the FCM and
WFM. The vertical axis is the multipole $\ell$ up to
$\ell=600$ and the horizontal the $m$ axis where $m \le \ell$. Due to the relation
$\alm=a^{*}_{\ell,-m}$, only modes from non-negative
$m$ are shown. Although the phase gradient (from neighboring modes) is the
most primitive way of qualitatively checking phase correlations, the
apparent presence of stripes shown in the FCM indicates strong coupling
between modes of neighboring $\ell$ of the same $m$.

\begin{apjemufigure}
\centerline{\includegraphics[width=0.99\linewidth]{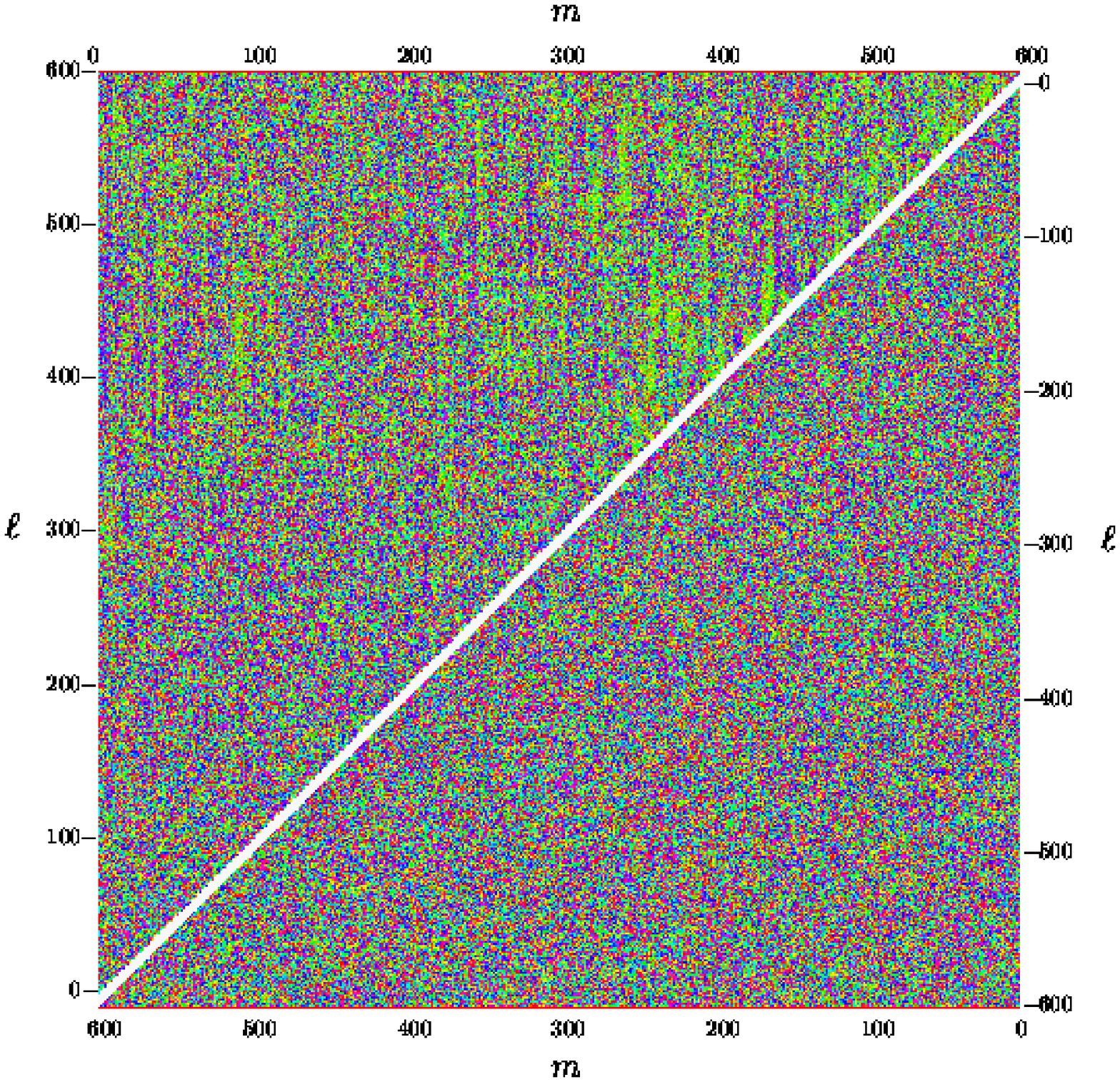}}
\caption{Color-coded phase gradient $D_\ell$ for the FCM (upper left
triangle) and
the WFM (bottom right). The vertical axis is the $\ell$ up to $\ell=600$ 
and horizontal the $m$ axis. Due to the relation
$a_{\ell,m}=a^{*}_{\ell,-m}$, we only show modes from non-negative
$m$. Although the phase gradient (from neighboring modes) is the most
primitive, the stripes shown from the FCM indicate strong phase
correlation between modes of neighboring $\ell$ of the same
$m$.}\label{color} 
\end{apjemufigure}

\section{Phase mapping and the mean chi-square statistic of the derived maps}
To test the Gaussianity of the FCM and the WFM based on the
random-phase hypothesis, we apply a phase mapping technique
\citep{mapping,chisquare} to quantify the degree of `randomness' of
the phases (i.e. Gaussian). The return map of phases is a bounded
square in which all phase pairs of fixed separation $(\Delta m, \Delta \ell)$ are mapped
as points (see Fig.~\ref{returnmap}). For example, one single return map for phase pairs with separation
$(\Delta m, \Delta \ell)=(0,1)$ contains points with $(x,y)$
coordinate $(\phi_{\ell,m}, \phi_{\ell+1,m})$, i.e. all phase pairs
from modes that are separated by $\Delta \ell=1$. If the phases are
random, we expect to have an ensemble of return maps of
all possible separations, each of which should be a scatter plot. As
such we are testing the `randomness' on the most strict terms. 
After mapping phase pairs on to a return map, we can apply a {\it
mean} $\chi^2$ statistic on the return map, which is defined as 
\begin{equation}
\meanchi=\frac{1}{M}\sum_{i,j}\frac{\big[ p(i,j)-\overline
p\big]^2}{\overline p},
\label{eq:meanchisquare}
\end{equation}
where $M$ is the number of pixels on the return map, $\overline p$ is the
mean value for each pixel on the discretized return
map. \citet{chisquare} have shown that for a homogeneous and isotropic GRF,
return mapping of phases results in an ensemble of return maps, each
with a Poisson distribution. The expectation value of the $\meanchi$
from such ensembles of Poisson-distributed maps is 
\begin{equation}
\brameanchi_{\rm P} = \frac{1}{4 \pi R^2}, \label{eq:theoreticalvalue}
\end{equation}    
where $R$ is the scale of smoothing from a 2D Gaussian convolution in order
to probe the spatial structure. The $\meanchi_{\rm P}$ will have a statistical
spreading around $ \brameanchi_{\rm P}$ with a dispersion $\Sigma_{\rm
P}$ where 
\begin{equation}
\Sigma_{\rm P}^2=\frac{1}{\pi^3 R^2 (M/2)}.
\end{equation}

\begin{apjemufigure}
\centering
\centerline{\includegraphics[width=0.95\linewidth]{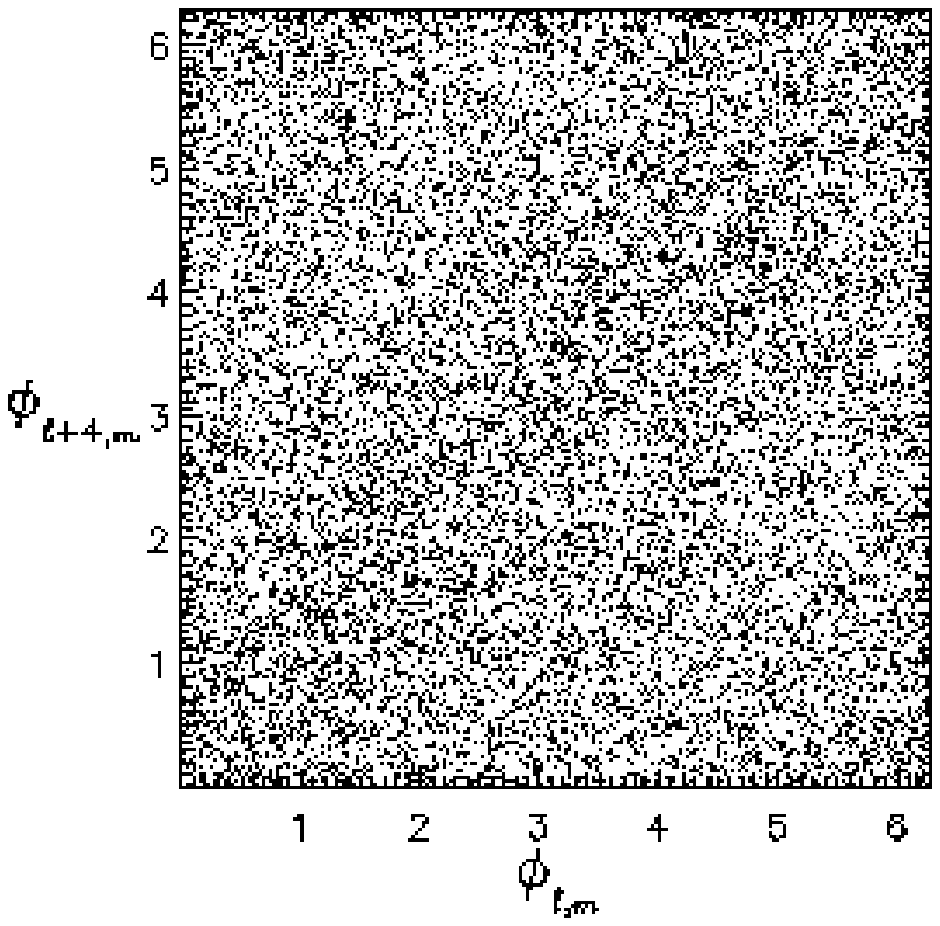}} 
\caption{An example of a return map for $(\Delta m, \Delta \ell)=(0,4)$
of phases $\phi_{\ell,m}$ of the FCM where $41 < \ell <250$. The
$\meanchi$ of this return map is 0.0332 when it is discretized into
$128^2$ pixels with smoothing scale $R=2$.} \label{returnmap}
\end{apjemufigure}

Figure~\ref{histograms} shows the histograms of the $\meanchi$
statistics from the ensemble of the return maps of the FCM and the
WFM for 4 multipole bands. One of the advantages of phase mapping
technique is that we are able to check Gaussianity in different
multipole bands, in particular those corresponding to foreground
contamination and noise. Here we present the $\meanchi$ statistic at 4
bands centered around $\ell \simeq 150$, 290, 400 and 500: $41 <\ell < 250$ (roughly the first
Doppler peak), $224 < \ell < 350$, $ 350 < \ell < 450$ and $ 463 <
\ell < 550$. The solid dark and dotted gray curves are the WFM and
FCM, respectively. In each panel the vertical line denotes the
expectation value  $\brameanchi_{\rm P} =(4 \pi R^2)^{-1}$. The curves
from the FCM are obviously skewed, and hence are manifestations of phase
correlations (i.e. non-Gaussian).

\begin{apjemufigure}
\centering
\centerline{\includegraphics[width=0.99\linewidth]{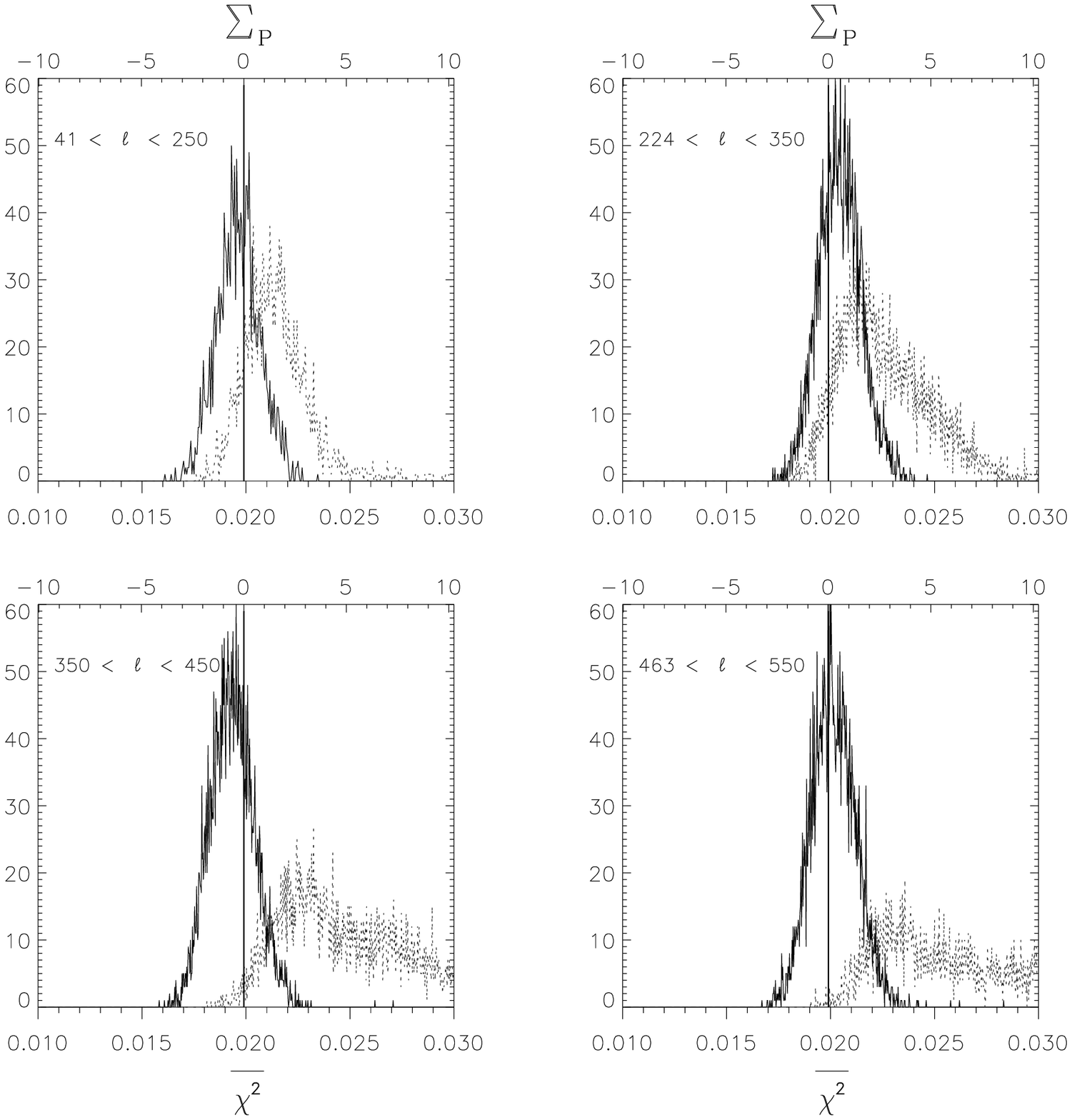}}
\caption{The histograms of $\meanchi$ statistic for the FCM (the
dotted gray curves) and the WFM (the solid dark curves) at different
multipole ranges $\ell$. One of the advantages of the phase mapping
technique is that it enables us to check non-Gaussianity for different
multipole ranges. The upper horizontal axis is annotated in terms of
the theoretical dispersion $\Sigma_{\rm P}$ of GRFs with origin set
at the expectation value $\brameanchi_{\rm P} =(4 \pi R^2)^{-1}$ (vertical
line in each panel). The smoothing scale on the $M=128^2$ discretized
return map is $R=2$.} 
\label{histograms}
\end{apjemufigure}

In Fig.~\ref{CL} we display the gross behavior of the distribution
curves in terms of the arithmetic mean $\brameanchi$ and the dispersion
$\Sigma$ from the mean chi-square statistic. The top panel is from the FCM and the bottom WFM. The contours mark 68\% (solid curve) and
95\% (dotted curve) CL regions from 2000 realizations
of GRFs. The symbols corresponds to 4 multipole bands centered at
$\ell \simeq 150$, 290, 400 and 500. Note that the contour region in the
bottom panel corresponds to a small section in the top panel. The phases of the 4 multipole bands from the FCM are
all strongly correlated, so they are far away from the 95\% CL
region. The WFM,
however, shows that phases below the first Doppler peak are random, 
with the other 3 multipole bands around the edge of 68\% CL region.

We see evidence of non-Gaussianity, however, in the WFM of the following two
bands centered $\ell \simeq 400$ and 500. In the lower two
panels of Fig.~\ref{histograms} there are points appearing at the
tails above $6\Sigma_{\rm P}$. On the other hand, among the 2000
realizations we simulate for GRFs, {\it no} mapping of phases reaches
$\meanchi$ value over $6 \Sigma_{\rm P}$, setting the probability
below 0.05\% for a GRF to have such mapping. Phase mapping from the separation
$(\Delta m, \Delta \ell)=(0,2)$ produces $\meanchi$ value at $7.3 
\Sigma_{\rm P}$ for the multipole band centered $\ell \simeq 400$,
also at $6.5 \Sigma_{\rm P}$ at $(\Delta m, \Delta \ell)=(1,2)$.  For
the band $\ell \simeq 500$, $7.6 \Sigma_{\rm P}$ appears at $(\Delta m,
\Delta \ell)=(2,2)$. These phase couplings are clear signs against the
random-phase hypothesis, therefore a manifestation of non-Gaussianity.

\begin{apjemufigure}
\centerline{\includegraphics[width=0.99\linewidth]{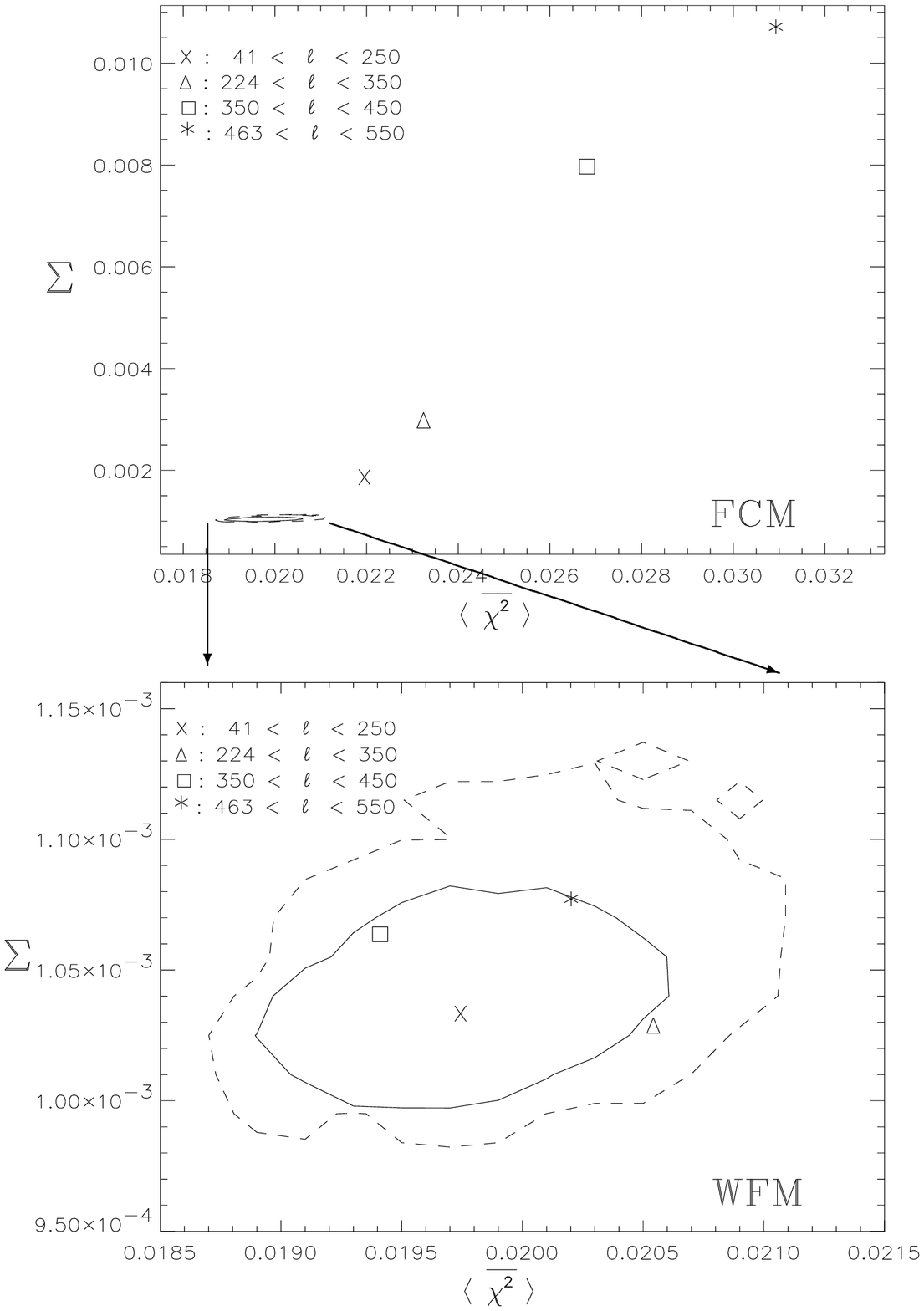}}
\caption{The mean chi-square statistic from the FCM (top) and WFM
(bottom) against
2000 realizations of GRFs, which is displayed in terms of the
arithmetic mean $\brameanchi$ and the dispersion $\Sigma$ of their
distribution curves. The contours mark 68\% (solid curve) and 95\%
(dotted curve) CL regions from 2000 realizations of
GRFs. Although 68\% and 95\% denotes 1 and 2-$\sigma$ deviation in Gaussian
statistics, the distribution is not {\it Gaussian} but rather
chi-square. The cross ($\times$), triangle
($\bigtriangleup$), square ($\Box$) and star ($\ast$) symbols
denote $\meanchi$ statistic from multipole ranges centered $\ell \simeq
150$, 290, 400 and 500, respectively. Note that the contour region in
the bottom panel corresponds to a small section in the top panel.}
\label{CL}
\end{apjemufigure}

We plot in Fig.~\ref{skymaps} the CMB temperature map from only two
multipoles $\ell=350$ and 352 (of all $m$'s) of the FCM and WFM. The choice of
these specific multipoles of $\Delta \ell=2$ from our previous
calculation is to demonstrate non-Gaussian signals the correlated phases will produce in the map. The structures at $\varphi \simeq 0$ and $\pi$ in the FCM, the residual signal after foreground cleaning, disappear after Wiener filtering. 

\begin{apjemufigure}
\centering
\centerline{\includegraphics[width=0.99\linewidth]{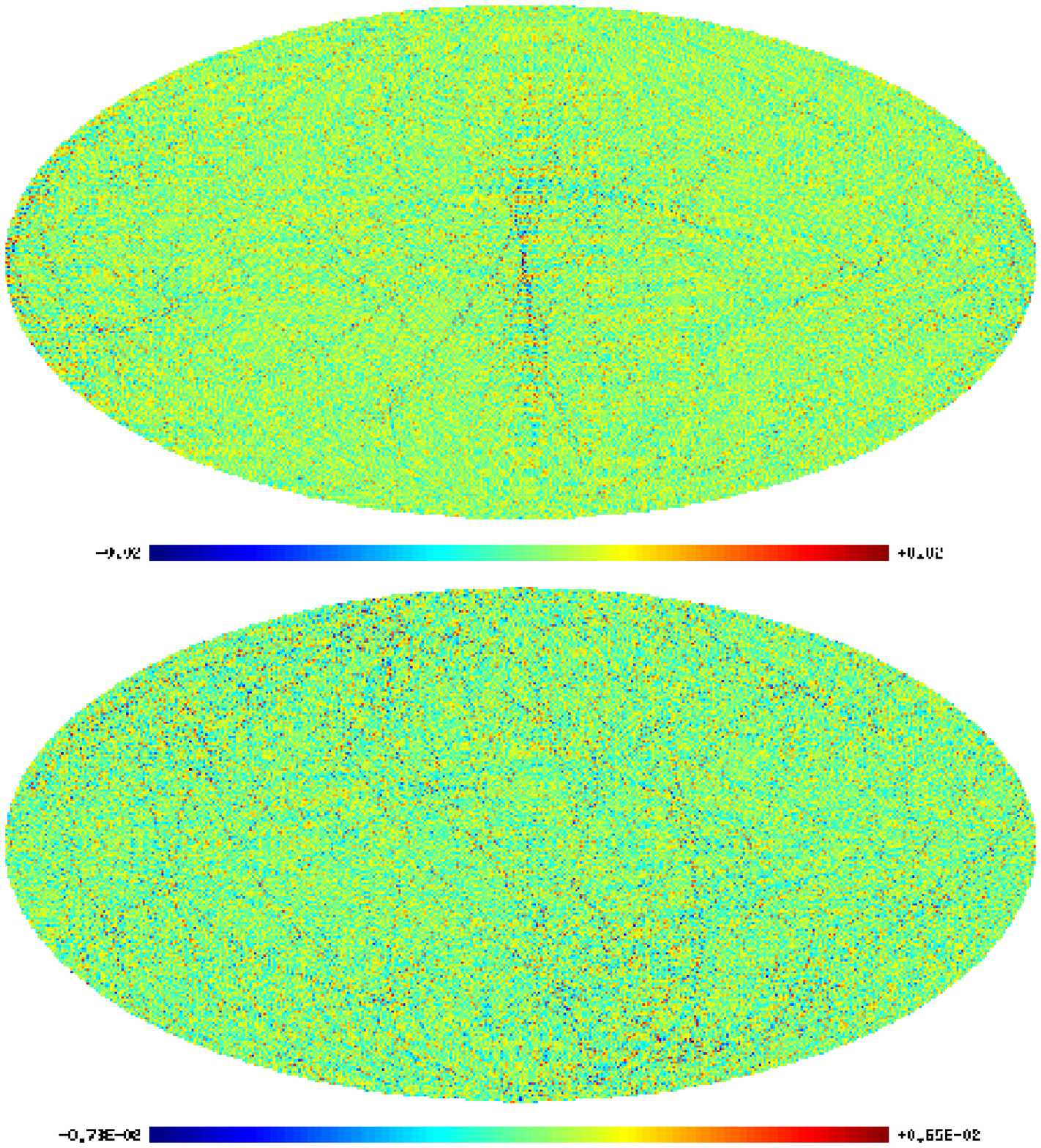}}
\caption{The CMB temperature from two multipoles $\ell=350$ plus 352 of the
FCM (top) and the WFM (bottom). These two
multipole modes are chosen because of the pronounced coupling between
modes $\Delta \ell =2$ of all $m$'s. The structures at $\varphi \simeq 0$ and $\pi$ shown in the FCM 
disappear after Wiener filtering, from which it is
marginally Gaussian at these two multipoles.} \label{skymaps}
\end{apjemufigure}
\section{Discussions}
In this {\it Letter} we have tested non-Gaussianity of two
maps: the Foreground-cleaned map and the Wiener-filtered map, which
are processed by TDH03 from the \wmap data. Based on the random-phase
hypothesis, we use a phase mapping technique to yield a statistic
that has detected considerable non-Gaussian signals for both maps at
most multipole bands. Our phase mapping technique is particularly
useful in separating non-Gaussian contributions from different sources
when various contaminations are present at different $\ell$ ranges. A
multipole band which is considerably non-Gaussian could have an
insignificant non-Gaussian contribution in the whole map and still produce
an over-all Gaussian realization within a certain confidence
level. As the uncertainties in foreground cleaning
propagate through the data-processing pipelines to the accuracy of the
angular power spectrum, it is therefore necessary to have effective
methods in component separation. We believe that our phase mapping
technique is a useful criterion to be incorporated into such
methods. Our statistic based on phase mapping also holds great advantage when
it comes to the issue of creating many whole-sky Gaussian realizations
for Gaussian statistics. As our null hypothesis is that phases are
random, we only need to put random phases (with Gaussian
instrumental noise being automatically included) for each harmonic
mode, which is easily done without any limit on the highest harmonic
number $\ell$ from any pixelization scheme. It is worth mentioning that
the upcoming \planck mission will have higher sensitivity and
resolution, hence every step of data processing will be crucial in
reaching such precision. 

\acknowledgements
This paper was supported by Danmarks Grundforskningsfond
through its support for the establishment of the Theoretical
Astrophysics Center. We thank Max Tegmark et al. for providing their processed
maps and making them public with openness. We thank Peter Coles and
Max Tegmark for useful discussions. We also acknowledge the use of \healpix
 package \citep{healpix} to produce $\alm$ and Fig.~\ref{skymaps}.

\end{document}